# Emergent Spin-Filter at the Interface between Ferromagnetic and Insulating Layered Oxides

Yaohua Liu[1], F. A. Cuellar[2], Z. Sefrioui[2], J. W. Freeland[3], M. R. Fitzsimmons[4], C. Leon[2], J. Santamaria[2] and S. G. E. te Velthuis[1]

[1]*Materials Science Division, Argonne National Laboratory, Argonne, Illinois 60439, USA.*
[2]*GFMC, Departamento de Fisica Aplicada III, Universidad Complutense de Madrid, Campus Moncloa, ES-28040 Madrid, Spain.*
[3]*Advanced Photon Source, Argonne National Laboratory, Argonne, Illinois 60439, USA.*
[4]*Los Alamos National Laboratory, Los Alamos, New Mexico 87545, USA.*

We report a strong effect of interface-induced magnetization on the transport properties of magnetic tunnel junctions consisting of ferromagnetic manganite $La_{0.7}Ca_{0.3}MnO_3$ and insulating cuprate $PrBa_2Cu_3O_7$. Contrary to the typically observed steady increase of the tunnel magnetoresistance with decreasing temperature, this system exhibits a sudden anomalous decrease at low temperatures. Interestingly, this anomalous behavior can be attributed to the competition between the positive spin polarization of the manganite contacts and the negative spin-filter effect from the interface-induced Cu magnetization.

PACS numbers: 75.25.-j, 73.40.Gk, 72.25.Mk, 75.70.Cn

Transition-metal oxide heterostructures are of keen interest because modified bonding at the epitaxial interfaces can give rise to fundamentally new phenomena and valuable functionalities [1-11]. The recently discovered interface-induced magnetization in several layered oxide systems has triggered increasing efforts to explore its influence on macroscopic properties [7-10]. Several studies have focused on how the interface-induced magnetization affects the magnetization reversal, in which exchange bias effects are observed [7,9], while its effects on the charge transport properties have been less studied. Of particular interest is the large uncompensated Cu moment at the interface between superconducting $YBa_2Cu_3O_{7-\delta}$ (YBCO) and ferromagnetic $La_{0.7}Ca_{0.3}MnO_3$ (LCMO), which arises from strong covalent bonding between the d-orbitals of Cu and Mn [5]. An effective ferromagnetic exchange field accompanies the interfacial Cu magnetization, extending into the cuprate layer and giving rise to the Jaccarino-Peter like magnetoresistance effect [10,12].

In this Letter we study the effects of the interfacial Cu magnetization on the transport properties of magnetic tunnel junctions (MTJs) consisting of an insulating $PrBa_2Cu_3O_7$ (PBCO) barrier [13] and ferromagnetic $La_{0.7}Ca_{0.3}MnO_3$ electrodes. The Ca- and Sr-doped $LaMnO_3$ families are renowned for half-metallicity–the conduction electrons at the Fermi surface are highly spin polarized at low temperatures [14-16]. The tunneling process is very sensitive to the delicate spin and electronic structures of both the metallic contacts and the insulating barrier [17,18], as well as the ferromagnet-insulator interface [19-21]. Therefore, the tunnel magnetoresistance *TMR* of MTJs is frequently employed to determine the spin polarization of ferromagnetic metals [15,22]. The TMR is computed using the Julliere formula, in which $TMR = \frac{G^P - G^{AP}}{G^{AP}} = \frac{2 p_1 p_2}{1 - p_1 p_2}$, where $G^P$ and $G^{AP}$ stand for the conductance of the parallel and antiparallel configurations of the two ferromagnetic electrodes, respectively, and $p_1$ and $p_2$ are the spin polarizations of the effective tunneling density of states in the two FM electrodes [23]. When temperature decreases, spin polarization usually increases, resulting in an enhancement of the TMR because the product of $p_1$ and $p_2$ approaches 1. However, for these cuprate-manganite MTJs, the *TMR* exhibits an anomalous and dramatic decrease, rather than the expected steady increase at low temperatures. Polarized neutron reflectometry (PNR) and x-ray magnetic circular dichroism (XMCD) studies on LCMO-PBCO-LCMO trilayers show that the saturation magnetization of the LCMO contacts increase as the temperature decreases. In other words, degradation of the ferromagnetic contacts is ruled out as a cause. Instead, we show that the anomalous temperature dependence is related to the interfacial Cu magnetization indicating that the spin degeneracy of the conduction band of the PBCO barrier is lifted and thus the barrier becomes spin selective. We conclude that the anomalous temperature dependence can be attributed to the competition between the positive spin polarization of the LCMO electrodes and the negative spin-filter effect from the interfacial Cu magnetization.

Trilayers with nominal structures of 8 nm LCMO /(2.4-7.2) nm PBCO /(25-50) nm LCMO were grown on (001)-oriented $SrTiO_3$ substrates via a high-$O_2$-pressure sputter deposition [24]. The trilayers were patterned into 16-100 μm² square-shaped MTJs for magnetotransport studies, using standard lithography and Ar ion milling techniques. The MTJs show high quality barriers free of defects or pinholes over large areas [24].

Magnetotransport experiments were conducted with current perpendicular to the sample plane via the four-terminal dc method. An in-plane magnetic field was applied along the [110] direction after cooling the junctions in a 4-kOe field. The magnetoresistance curves of these junctions displayed abrupt resistance switching between parallel (low resistance) and antiparallel (high resistance) states (See Fig. S1). The onset temperature of TMR is limited by the Curie temperature of the top LCMO contact of each junction. Note that the abrupt switching between parallel and antiparallel magnetization states of the electrodes is typical of magnetic tunnel junctions and is usually not found in other forms of (hopping) transport in manganites. Figure 1(a) shows the

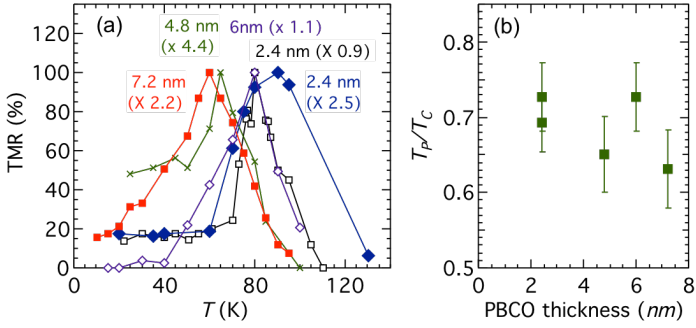

FIG. 1 (color online). (a) Temperature dependence of the TMR of the MTJs consisting of $La_{0.7}Ca_{0.3}MnO_3$ electrodes and $PrBa_2Cu_3O_7$ barriers wth different barrier thicknesses, exhibiting an anomalous suppression in TMR at low temperatures. The TMR is scaled for different junctions for comparison. (b) The reduced TMR peak temperature, with respect to the TMR onset temperature, slightly decreases as the PBCO thickness increases. The error bar comes from the uncertainty of the onset temperature, which is ~ 5 K.

temperature dependence of the TMR for five junctions. Counter-intuitively, the TMR amplitude displays an increase but then a surprising decrease on cooling. The TMR peak temperature is labeled as $T_P$. The decrease of TMR at low temperatures is very anomalous and to our knowledge has not been reported in literature on layered-oxide based MTJs. Both the maximum TMR and $T_p$ vary from junction to junction, with $T_P$ changing between 60 K and 90 K. The ratio between the $T_p$ and the TMR onset temperature, $T_C$, slightly changes between 0.63 and 0.73 among different junctions [Fig. 1(b)]. As temperature decreases, the spin polarization of the ferromagnetic manganite normally increases [22] and the spin-flip scattering decreases [23,25], both of which will enhance TMR as usually reported [15,22]. Therefore, the observed TMR suppression at low temperatures is unusual.

We performed PNR experiments to characterize the magnetization of the LCMO electrodes, using the Asterix reflectometer at the Lujan Neutron Scattering Center of Los Alamos National Laboratory. PNR is capable of resolving the depth profile of magnetization with nanometer resolution [26]. The sample was cooled to 12 K in a 5 kOe in-plane field along [110] direction. Subsequently, data were collected at 12 K and then at 80 K in saturation (H = 5 kOe). We also measured x-ray reflectivity (XRR, data not shown) at room temperature. XRR and PNR data were model-fitted using the Parratt formalism to optimize depth dependent scattering length density (SLD) profiles [27]. The PNR data and the best fit are displayed in Fig. 2(a). Figure 2 (b) shows the depth profiles of the neutron nuclear scattering length density as well as the saturation magnetization at 12 K and 80 K inferred from the data. PNR lacks the chemical specificity of resonant soft x-ray experiments, the latter of which are used to resolve the interfacial Cu moment as discussed below. The PNR results show that in fact the magnetic SLD profile follows the structural SLD profile at the barrier interfaces, hence there is no evidence of a magnetic "dead" layer at the interfaces. The saturation magnetization of both the top and

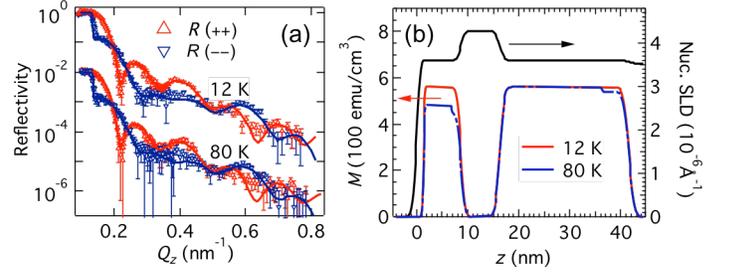

FIG. 2 (color online). (a) Polarized neutron reflectivity in saturation (H = 5 kOe) at 12 K and 80 K. The best-fit curve (line) is overlaid on the data (circles). (b) Depth profiles of the neutron nuclear scattering length density and the saturation magnetization at 12 K and 80 K.

bottom LCMO layers were found to be 3.6 $\mu_B$ per Mn ion at 12 K, which closely matches the value of the optimally doped (La, Ca)$MnO_3$ for the half-metallic phase. Therefore, both the top and bottom LCMO electrodes are expected to have a high spin polarization near 100% at 12 K.

Figure 2(b) shows that the saturation magnetization of the bottom LCMO layer is almost same at 12 K and 80 K, however the saturation magnetization of the top LCMO layer is significantly less at 80 K compared to 12 K. To further explore the temperature dependence of the magnetization, we used XMCD and SQUID magnetometery (Fig. 3). The XMCD experiments were conducted at the beamline 4-IDC at Advanced Photon Source. Circularly polarized X-rays were used to obtain absorption spectra recorded by total electron yield (TEY) at a grazing x-ray incidence angle of 10°. The XMCD spectra are given by the difference between the absorption spectra of the right and left circularly polarized x-rays, normalized by the peak jump at the $L_3$ edge of the average absorption spectra. The data were collected in the remnant states after saturation in both positive and negative 1 kOe in-plane field along the [100] direction to rule out experimental artifacts. Figure 3 shows the temperature dependence of the XMCD peak values at the $L_3$ edges of Mn and Cu, respectively, reflecting the amplitude of the element specific magnetization. We have confirmed that the interfacial Cu magnetization is antiparallel to the Mn magnetization from XMCD. Furthermore, we have confirmed that the net Cu moment exists at both the top and bottom LCMO/PBCO interfaces with resonant magnetic x-ray scattering (RMXS) (data not shown), suggesting a symmetrical interface structure as previously observed in LCMO/YBCO heterostructures [28]. The temperature dependence of the Mn XMCD signal is well described by the empirical formula for spontaneous magnetization, $M(T) = M(0)\left[1 - \left(\frac{T}{T_C}\right)^\alpha\right]^\beta$ with the critical exponent $\alpha = 1.5$ from Bloch's law [22], and $T_C = 141$ K and $\beta = 0.5$ from the fitting. The XMCD spectra are primarily sensitive to the top LCMO layer but not the bottom one because of the limited electron escape depth of the TEY signal (~ 3-5 nm); while the SQUID magnetometer measures the magnetization of the whole sample. Therefore, the XMCD data indicate that the

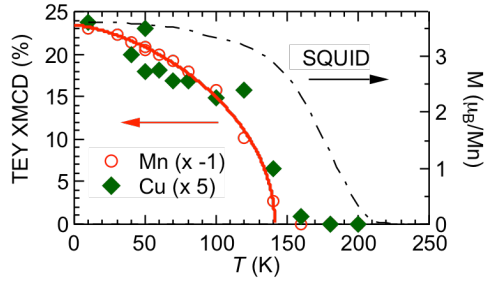

FIG. 3 (color online). Temperature dependences of the XMCD peak intensities at Mn $L_3$-edge (circles at 641.8 eV) and Cu $L_3$-edge (diamonds at 947.8 eV), respectively. The data were collected in the remnant states after applying a 1 kOe in-plane field along the [100] direction. The solid line is the best fit to the Mn signal using the modified Bloch law. Also shown is the magnetization acquired with a SQUID magnetometer during cooling in a 100 Oe field.

Curie temperature ($T_C$) of the top LCMO layers is about 140 K, while the SQUID data show that the bottom LCMO layer has a higher $T_C$ of 200 K. The much-decreased $T_C$ of the top LCMO layer explains why its saturation magnetization clearly decreases when temperature increases from 12 K to 80 K. Overall, PNR and XMCD studies confirm that there is no degradation of the magnetization of the LCMO electrodes at low temperatures.

The observed interfacial Cu magnetization indicates that the spin degeneracy of the conduction band of the PBCO barrier is lifted. A tunnel junction with an exchange-split barrier may displayed complex transport phenomena, which can be attributed to the so-called spin-filter effect [29-31]. Below we argue that a 'negative' spin-filter effect accompanying the interfacial Cu magnetization strongly affects the spin dependent tunneling process. The effect is illustrated in Fig. 4(a) through evaluating the spin polarization inside the LCMO FM electrode ($p$) and the effective spin polarization inside the PBCO barrier ($p^{eff}$). When the kinetic energy of an electron is less than the barrier height, the wave function exponentially decays across the barrier. The decay rate, within in the Wentzel-Kramers-Brillouin (WKB) approximation [32], depends on the height and width of the barrier. In the case of a spin-independent barrier, the wave functions of spin-up and spin-down electrons have the same decay rate and $p^{eff} = p$. The situation changes for a spin-dependent barrier. Since the induced Cu net moment is antiparallel to the Mn magnetization of the adjacent layer, the interfacial PBCO layer is less transparent to the majority-spin electrons of the adjacent LCMO electrodes. Therefore, the wave function of the majority-spin electrons decays faster when they penetrate into the barrier than the minority-spin electrons. Note that the optimal doped manganite has a positive spin polarization $p$ [16]. Thus $p^{eff}$ is less than $p$, which we call the negative spin-filter effect.

To quantify this behavior, we consider an interfacial PBCO region with an exchange splitting of $2\Delta_{ex}$ in the conduction band. After tunneling through this interfacial region, $p^{eff} = \frac{(1+p)T_\uparrow - (1-p)T_\downarrow}{(1+p)T_\uparrow + (1-p)T_\downarrow}$ with $T_{\uparrow,\downarrow} = exp\left\{-2\int_0^d \sqrt{\frac{2m^*}{\hbar^2}(\Phi_0 \pm \Delta_{ex})}\,dx\right\}$, where $d$ is the effective width of the exchange-split region, $m^*$ is the transport effective mass and $\Phi_0$ is the average barrier height. $T_{\uparrow,\downarrow}$ reflects the different decay rates for the wave functions of the spin-up and spin-down electrons inside the exchange-split region [Fig. 4(a)]. The spin polarization of the LCMO electrode $p$ and the exchange splitting in the interfacial region $\Delta_{ex}$ are correlated in a subtle way. First, $\Delta_{ex}$ is proportional to the Cu magnetization. Second, the spin polarization and the magnetization of $La_{2/3}Sr_{1/3}MnO_3$ have similar temperature dependencies [22], which we assume is the case for LCMO

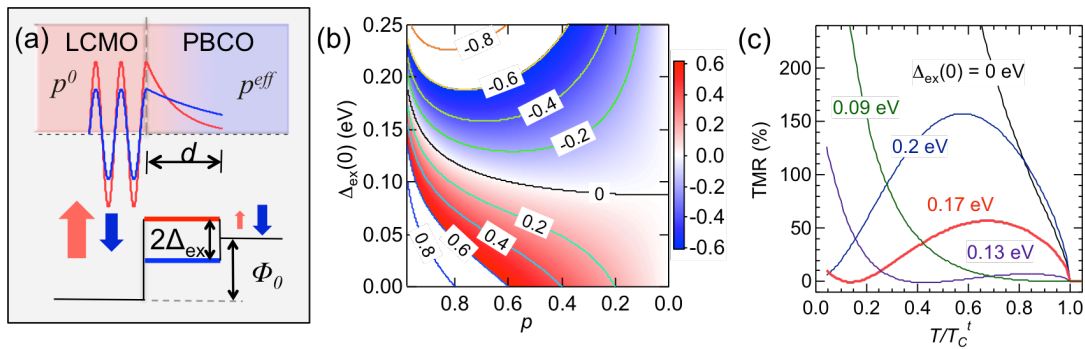

FIG. 4 (color online). (a) The interface Cu magnetization gives rise to a negative spin-filter effect due to lifted spin degeneracy, thus the majority-spin electrons (spin up, labled by red color) of the LCMO electrode experience a higher barrier than the minority-spin electrons (spin down, labed by blue color) when tunneling through the interfacial PBCO region. The lines in the top panel illustrate the spin-dependent wavefunctions of the charge carriers. In the bottom panel, the colored thick lines illustrate the spin-dependent conduction bands in the interfacial PBCO region, and the size of the arrows shows the populations of the spin-up (spin-down) charge carriers. (b) Effective spin polarization as a function of the spin polarization of the LCMO electrode $p$ and the zero-temperature exchange splitting $\Delta_{ex}(0)$ in the interfacial PBCO region. (c) Calculated TMR as a function of the reduced temperature for different $\Delta_{ex}(0)$. The TMR displays complex temperature dependences for high $\Delta_{ex}(0)$.

electrodes. Third, the Cu magnetization tracks the Mn magnetization when the temperature changes (Fig. 3 and Ref.[4]). Therefore, both $p$ and $\Delta_{ex}$ have a similar temperature dependencies to the magnetization of the LCMO electrode. Although both $p$ and $\Delta_{ex}$ increase as the temperature decreases, they have opposite influences on $p^{eff}$. Their mutual competition may result in complex temperature dependencies of $p^{eff}$ and consequently of the TMR.

We first calculate $p^{eff}$ as a function of $p$ and the zero-temperature exchange splitting $\Delta_{ex}(0)$. As discussed above, at a certain temperature, the exchange splitting $\Delta_{ex}$ has the amplitude of $\frac{p}{p(0)}\Delta_{ex}(0)$, where $p(0)$ is the zero-temperature spin polarization of the LCMO electrode, which has been considered in the calculations. The effective width ($d$) of the exchange-split region is assumed to be one unit cell (1.2 nm) of PBCO along the tunneling direction (c-axis) because the induced Cu magnetization is localized at the interface [12]. The average barrier height of 0.3 eV is estimated from the *I-V* curves of the junctions using the Brinkman-Rowell-Dynes (BDR) tunneling formula [33]. Figure 4 (b) shows the calculated $p^{eff}$ using $d = 1.2$ nm, $\Phi_0 = 0.3$ eV, $p(0) = 0.98$, and $m^* = m_e$, where $m_e$ is the free electron mass. As expected, $p^{eff}$ decreases as $\Delta_{ex}(0)$ increases. For a sufficiently large $\Delta_{ex}(0)$, the $p^{eff}$ becomes negative and shows a non-monotonic dependence on the spin polarization (thus the temperature) with the maximum amplitude of $p^{eff}$ occurring approximately at $p = \left[1 - \frac{\hbar p(0)}{d\Delta_{ex}(0)}\sqrt{\frac{\Phi_0}{2m^*}}\right]^{1/2}$, provided that $\Delta_{ex}(0) < \Phi_0$.

The TMR at an infinitesimal bias can be calculated in the WKB approximation with a square barrier model, which is expressed in the Julliere formula using $p^{eff}$ to include the effect of the interfacial Cu magnetization. As discussed above, the temperature dependencies of $p$ and $\Delta_{ex}$ follow the temperature dependence of the LCMO magnetization. Thus they are formulated in the following forms: $p(T) = p(0)\left[1 - \left(\frac{T}{T_c}\right)^{\frac{3}{2}}\right]^{0.5}$ and $\Delta_{ex}(T) = \Delta_{ex}(0)\left[1 - \left(\frac{T}{T_c}\right)^{\frac{3}{2}}\right]^{0.5}$. Without loss of generality, we assume that the Curie temperatures of the top ($T_C^t$) and bottom ($T_C^b$) LCMO contact are different where $T_C^b = 1.4 T_C^t$, and both have zero-temperature spin polarizations $p(0)$ of 98%. Figure 4(c) shows the calculated TMR as a function of the reduced temperature $t = T/T_C^t$ for different $\Delta_{ex}(0)$. The amplitude of $\Delta_{ex}(0)$ reflects the strength of the Cu-Mn covalent bond at the interface. For $\Delta_{ex}(0) = 0$ eV, the TMR increases sharply when the temperature decreases. For $\Delta_{ex}(0) = 0.09$ eV (the weak covalent-bond case), TMR becomes lower due to a decreased $p^{eff}$, but the TMR still increases steadily as the temperature decreases. Interestingly, when $\Delta_{ex}(0)$ is above 0.13 eV, the temperature dependence evolves into a complex behavior. As $\Delta_{ex}(0)$ further increases, the complex dependence becomes much pronounced, and the TMR peak temperature shifts to lower temperatures. When $\Delta_{ex}(0) = 0.17$ eV, $T_P$ reaches $0.68 T_C^t$, close to the observed reduced TMR peak temperature [Fig. 1(b)]. It is worth noting that the amplitude of $p^{eff}$ can be larger than $p$ for a large $\Delta_{ex}(0)$; therefore the TMR in the high $\Delta_{ex}(0)$ limit (*e.g.*, $\Delta_{ex}(0) = 0.20$ eV) can exceed the value in the case of zero $\Delta_{ex}(0)$, which can be seen in the temperature region close to $T_C^t$. We have further examined the effects on the TMR from the average barrier height $\Phi_0$, the effective width of the exchange-split region $d$, the possible enhanced effective mass $m^*$ and the decreased zero-temperature spin polarization $p(0)$ (Fig. S2). The non-monotonic temperature dependence exists for a broad parameter space. Approximately, the $\Delta_{ex}(0)$ for the TMR peak temperatures occurring at $0.68 T_C^t$ follows the criteria, $\Delta_{ex}(0) = \frac{\hbar p(0)}{d(1-0.44p(0)^2)}\sqrt{\frac{\Phi_0}{2m^*}}$. We note that the calculations do not precisely reproduce the TMR amplitude and the shape of the temperature dependence as experimentally observed. First-principle calculations incorporating the band structures at the interfaces may be able to resolve the discrepancies [19,34,35], which is beyond the scope of this work. However, the key feature has been reproduced within the WKB approximation, which clearly shows that the competition between the positive spin polarization of LCMO and the negative spin-filter effect from the interfacial Cu magnetization can give rise to the complex TMR temperature dependence.

Finally, we discuss the variation among different junctions (Fig. 1). Calculations show that the maximum TMR is highly sensitive, but the reduced TMR peak temperature is less sensitive to the amplitude of the exchange splitting in the interfacial PBCO region. Therefore, the large variation in the maximum TMR can be attributed to subtle differences in the Cu-Mn covalent strength among different samples. Provided that the interface condition is not significantly changed, the model predicts a lower TMR peak temperature with decreasing Curie temperature of the top LCMO, which agrees with the experiments [Fig. 1(a)]. Additionally, Fig. 1(b) shows that the reduced peak temperature only slightly decreases with increasing PBCO thickness, suggesting that the zero-temperature spin polarization of the LCMO electrodes barely depends on the PBCO barrier thickness.

In summary, MTJs consisting of ferromagnetic $La_{0.7}Ca_{0.3}MnO_3$ and insulating $PrBa_2Cu_3O_7$ show an anomalous TMR suppression at low temperatures. Our calculations, within the Wentzel-Kramers-Brillouin approximation, show that complex temperature dependence can arise from a competition between the high positive spin polarization in the manganite electrodes and a negative spin-filter effect from the interfacial Cu magnetization. This work illustrates that the recently discovered interface-induced magnetization in layered oxide heterostructures can have non-trivial effects on the macroscopic transport properties. The emergent interfacial magnetization appears common [4,6-9] and tunable [36,37], which thus provides many opportunities to engineer oxide spintroincs with tailored properties.

Work at Argonne National Laboratory and use of the Advanced Photon Source, was supported by the U.S.

Department of Energy, Office of Basic Energy Sciences under contract no. DE-AC02-06CH11357. Work at UCM was supported by Spanish MICINN through grant MAT2011-27470-C02, Consolider Ingenio 2010 - CSD2009-00013 (Imagine) and by CAM through grant S2009/MAT-1756 (Phama). This work has benefited from the use of the Lujan Neutron Scattering Center at LANSCE, which is funded by the Department of Energy's Office of Basic Energy Sciences. Los Alamos National Laboratory is operated by Los Alamos National Security LLC under DOE through Contract No. DE-AC52-06NA25396.

**Emergent Spin-Filter at the Interface between Ferromagnetic and Insulating Layered Oxides**

Yaohua Liu[1], F. A. Cuellar[2], Z. Sefrioui[2], J. W. Freeland[3], M. R. Fitzsimmons[4], C. Leon[2], J. Santamaria[2] and S. G. E. te Velthuis[1]

[1]*Materials Science Division, Argonne National Laboratory, Argonne, Illinois 60439, USA.*
[2]*GFMC, Departamento de Fisica Aplicada III, Universidad Complutense de Madrid, Campus Moncloa, ES-28040 Madrid, Spain.*
[3]*Advanced Photon Source, Argonne National Laboratory, Argonne, Illinois 60439, USA.*
[4]*Los Alamos National Laboratory, Los Alamos, New Mexico 87545, USA.*

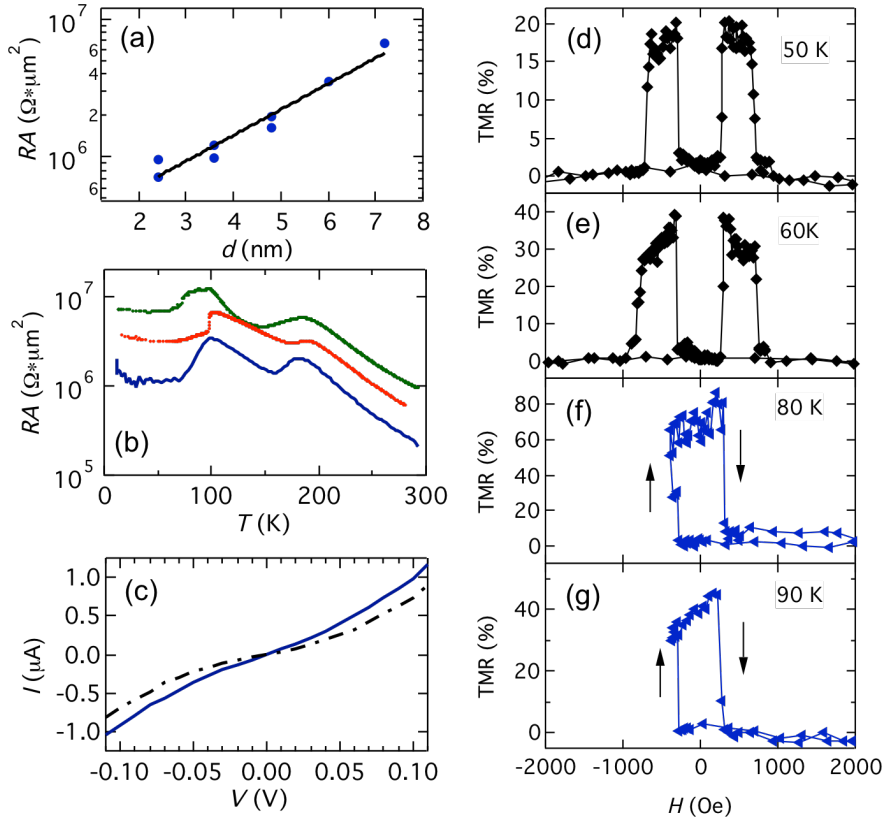

Figure S1. Tunneling properties of the junctions. (a) Resistance area product RA as a function of the PBCO barrier thickness under a bias of 10 mV and at 30 K in 4 kOe field. (b) Temperature dependence of RA from junction s with 7.2 nm, 6 nm and 2.4 nm PBCO barriers, from top to bottom, respectively. Data were collected in 4 kOe field. Noting the insulating behavior at low temperatures. (c-g) Transport data from a junction with 6 nm PBCO barrier. (c) Current as a function of voltage at 80 K in parallel (solid line) and antiparallel (dashed line) states. (d)-(g) TMR *vs.* field at different temperatures, including two full hysteresis loops and two minor loops.

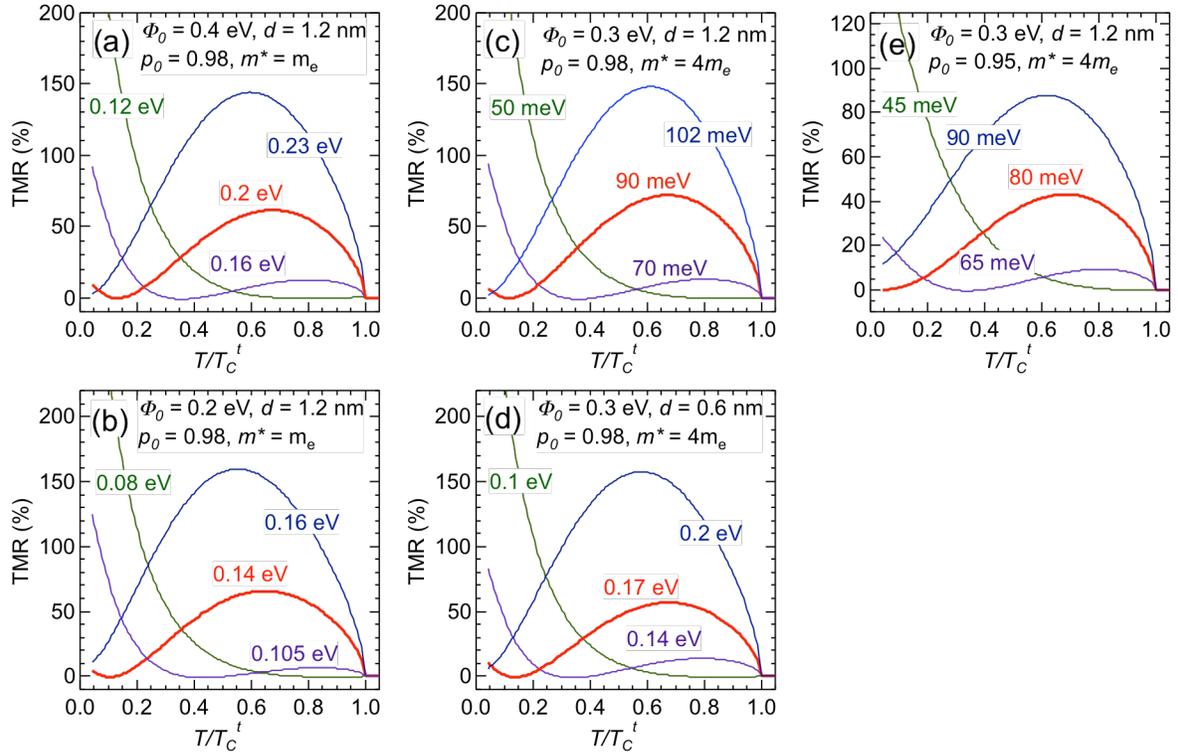

Figure S2. Calculated TMR as a function of the reduced temperature for different $\Delta_{ex}(0)$, similar to Fig. 4(c), but calculated with different values of [(a) and (b)] the average barrier height $\Phi_0$, (c) the possible enhanced effective mass $m^*$, and (d) the effective width of the exchange-split region $d$, and (e) the zero-temperature spin polarization $p(0)$.